\pdfoutput=1

\documentclass[9pt,twocolumn,twoside]{osajnl}

\journal{josaa} 

\setboolean{shortarticle}{false}

\usepackage[normalem]{ulem}
\usepackage{cancel}
\ifthenelse{\boolean{shortarticle}}{\colorlet{color2}{color2b}}{\colorlet{color2}{color2}} 

\title{Characterization of Power Absorption Response of Periodic 3D Structures to Partially Coherent Fields}

\author[1*]{Denis Tihon}
\author[2]{Stafford Withington}
\author[2]{Christopher N. Thomas}
\author[1]{Christophe Craeye}

\affil[1]{Universit\'{e} catholique de Louvain, ICTEAM Institute, Place du Levant 3, 1348 Louvain-la-Neuve, Belgium}
\affil[2]{Cambridge University, Cavendish Laboratory, J.J. Thomson Avenue, CB3 OHE Cambridge, UK}

\affil[*]{Corresponding author: denis.tihon@uclouvain.be}

\dates{Compiled \today}

\ociscodes{(000.4430) Numerical approximation and analysis; (010.5620) Radiative transfer; (030.4070) Modes; (050.1755) Computational electromagnetic methods; (300.1030) Absorption.}

\doi{\url{http://dx.doi.org/10.1364/ao.XX.XXXXXX}}

\begin{abstract}
In many applications of absorbing structures it is important to understand their spatial response to incident fields, for example in thermal solar panels, bolometric imaging and controlling radiative heat transfer. In practice, the illuminating field often originates from thermal sources and is only spatially partially coherent when reaching the absorbing device. In this paper, we present a method to fully characterize the way a structure can absorb such partially coherent fields. The method is presented for any 3D material and accounts for the partial coherence and partial polarization of the incident light. This characterization can be achieved numerically using simulation results or experimentally using the Energy Absorption Interferometry (EAI) that has been described previously in the literature. The absorbing structure is characterized through a set of absorbing functions, onto which any partially coherent field can be projected. This set is compact for any structure of finite extent and the absorbing function discrete for periodic structures. 
\end{abstract}

\setboolean{displaycopyright}{true}

\begin{document}

\maketitle
\thispagestyle{fancy}

\ifthenelse{\boolean{shortarticle}}{\ifthenelse{\boolean{singlecolumn}}{\abscontentformatted}{\abscontent}}{}


\section{Introduction}

Electromagnetic metamaterials with a range of behaviours not found in nature can be realised from 3D periodic structures where some or all of the dimensions of the unit-cell are smaller than the wavelength of the exciting field.  Metamaterials engineered in this manner are expected to provide a new means of subwavelength imaging \citep{intro1, intro2}, improved broadband absorbers \citep{intro3, 5} and to allow controlled radiative heat transfer between surfaces \citep{Simovski2013}, among many other uses.  In many of the envisaged applications, particularly those at higher frequencies (THz), the incident radiation is thermal in origin and so is only expected to be partially spatially coherent and in a partial state of polarization on reaching the structure. Examples include absorbers for thermal solar panels and bolometric detectors, as well as coatings for controlled radiative cooling.
Being able to model and characterize the response of 3D structures illuminated by partially coherent fields is therefore crucial for optimizing the performance of the metamaterials used. However, it is an area that has received little attention compared with understanding the response to fully coherent fields and frequency-dependent behaviour.

In a previous paper \citep{4}, a way of characterizing the absorption of fields in any state of coherence by a structure has been developed. A formal link between theoretical results and experimental data that can be collected using Energy Absorption Interferometry (EAI) \citep{3} has been derived. However, for the sake of clarity, only the 2D case and scalar wave function have been considered in  \citep{4}, with an extension to structures periodic along one direction. In this paper, we present a formalism for power absorption by arbitrary 3D structures that can take into account the partial coherence and polarization of any stationary incident fields, with a special treatment for the case of doubly periodic structures. We prove that it is possible to deal with the partially coherent nature of the excitation, either by decomposing the incident fields into a sum of fully coherent fields \citep{1,6} or by computing the set of natural absorption modes of the structure, which is independent from the excitation.

The opportunities of EAI are very large. In fact, the method is able to recover the natural dynamical modes of any energy absorbing structure. In the case of microwave and optical photon-counting detectors for quantum communications it is essential to avoid, or at least terminate carefully, electromagnetic modes that can only couple noise and stray light into the detector. In the case of energy harvesting components, antenna arrays and absorbers, it is essential to maximize the number of degrees of freedom available for collecting power. The same considerations apply to near-field energy and information transfer between separated or overlapping volumes.  In a recent paper \citep{answer2}, the method has been extended to recovering the collective excitations of quantum-mechanical systems.

This paper is organized as follows. In Section 2, we consider the problem of absorption of partially coherent fields by arbitrary structures and the treatment of the structure is decoupled from the treatment of the fields. In Section 3, the decomposition of the partially coherent fields into a sum of coherent modes is presented. In Section 4, the Energy Absorption Interferometry (EAI) technique is described from a theoretical point of view and, in Section 5, we derive a relation between the data collected and the characterization of the structure studied. In Section 6, the case of structures periodic in two directions is studied and a simplified formulation is proposed. In Section 7, we discuss the convergence of the infinite summations and integrations that are involved in the formulation. The computation of the absorption modes is presented and commented on in Section 8. A summary of the whole method is presented in Section 9 and last, in Section 10, numerical results are presented and commented upon.

\section{Power absorbed by 3D structures}
\label{sec:2}
Let us assume a 3D structure with any distribution of real permittivity ($\varepsilon$), permeability ($\mu$) and resistivity ($\rho$). The resistivity accounts for all the loss mechanisms, including dielectric losses. For clarity, magnetic losses are not considered hereafter. However, the extension of the proposed method to such losses is straightforward. 
We will assume that the incident fields are temporally stationary, as is this generally the case for fields of thermal origin.
For a stationary field, the cross-correlation function vanishes over time \citep{2}, which is equivalent to saying that the different frequency components of the field are fully incoherent.
As a result, the total power absorbed is simply the sum of the power absorbed at each of the different frequencies present, which we may consider separately.
$\exp(j \omega t)$ time dependence of the fields and currents will be implicitly assumed hereafter, with $\omega$ the angular frequency for which the structure is characterized. 

Let $\mathbf{E}_i(x,y)$ denote the incident electric field over the reference plane $z=0$ located above the structure. The basis formed using the $x$, $y$ and $z$ axes is orthonormal and right-handed.
For later convenience, we decompose the incident field into vector components resulting from Transverse-Electric (TE) and Transverse-Magnetic (TM) field modes:
\begin{equation}
\mathbf{\tilde{E}}_i(\mathbf{k}_t) = 
\begin{pmatrix}
\tilde{E}_{i,\text{TE}}(\mathbf{k}_t) \\ \tilde{E}_{i,\text{TM}} (\mathbf{k}_t)
\end{pmatrix},
\end{equation}
with $\tilde{E}_{i,\text{TE}}(\mathbf{k}_t)$ or $\tilde{E}_{i,\text{TM}} (\mathbf{k}_t)$ the amplitude of the plane wave whose real transverse wave-vector is $\mathbf{k}_t = (k_x, k_y, 0)$ and whose electric or magnetic fields are transverse to the plane $z=0$, respectively. The polarization vectors of these plane waves are
\begin{subequations}
\label{eq:08-07-01}
\begin{align}
\mathbf{E}^{\text{TE}}(\mathbf{k}_t) &= \dfrac{1}{k_t} \big(-k_y \mathbf{\hat{x}}+k_x \mathbf{\hat{y}}\big) \\
\mathbf{E}^{\text{TM}}(\mathbf{k}_t) &= \dfrac{k_z}{k_t
k_0 
} \big(k_x \mathbf{\hat{x}} + k_y \mathbf{\hat{y}} + k_t^2/k_z \mathbf{\hat{z}} \big) ,
\end{align}
\end{subequations}
with $\mathbf{k} = (k_x, k_y, -k_z)$ the wave vector of the plane wave, $k_t = |\mathbf{k}_t|$ is the magnitude of the transverse wave-vector, $k_0 = \omega \sqrt{\varepsilon_0\mu_0}$ the free-space wave number and $k_z = \pm \sqrt{(k_0^2-k_t^2)}$ the propagation constant of the plane-wave along the $\mathbf{\hat{z}}$ direction. The plus sign has to be chosen for $k_t < k_0$ and the minus sign otherwise (i.e. $\arg(k_z) \in ~ \{-\pi/2, 0\})$. $\varepsilon_0$ and $\mu_0$ are the free space permittivity and permeability respectively.

The transformation from the spatial to the spectral representations takes the form
\begin{subequations}
\label{eq:18-09-01}
\begin{align}
\tilde{E}_{i,\text{TE}}(\mathbf{k}_t) &= \iint_{-\infty}^{\infty} \mathbf{E}_i(x,y) \cdot  
\mathbf{E}^{\text{TE}}(\mathbf{k}_t) 
 \exp\big(j \mathbf{k}_t \cdot \mathbf{r} \big) d\mathbf{r}  \\
\tilde{E}_{i,\text{TM}}(\mathbf{k}_t) &= \iint_{-\infty}^{\infty} \mathbf{E}_i(x,y) \cdot 
\mathbf{E}^{\text{TM}}(\mathbf{k}_t) 
 \exp\big(j \mathbf{k}_t \cdot \mathbf{r} \big) d\mathbf{r} ,
\end{align}
\end{subequations}

with $\mathbf{r} = (x,y,z)$ the position where the incident fields are estimated and $E^*$ the complex conjugate of $E$. The integration is carried out over the reference plane $z=0$. The dot product used throughout this paper is defined by $\mathbf{a}\cdot\mathbf{b}=\sum a_i b_i$.
In the rest of this paper, integration from $-\infty$ to $\infty$ should be assumed unless other limits are stated. 
The reverse operation, which consists of computing the spatial incident fields distribution in the plane $z=0$ using their plane wave spectrum, is given by
\begin{equation}
\begin{split}
\mathbf{E}_i(x,y) =  &\dfrac{1}{4\pi^2}\iint \Big(\tilde{E}_{i,\text{TE}}(\mathbf{k}_t) \mathbf{E}^{\text{TE}}(\mathbf{k}_t) \\
&+ \tilde{E}_{i,\text{TM}}(\mathbf{k}_t) \mathbf{E}^{\text{TM}}(\mathbf{k}_t) \Big) \exp\big(-j \mathbf{k}_t \cdot \mathbf{r} \big) d\mathbf{k}_t.
\end{split}
\end{equation}

The incident fields are partially coherent, which means that the total fields within the structure correspond to stochastic functions \citep{2}. 
The mean power density dissipated over time at point $\mathbf{r}$ can be computed using
\begin{equation}
\label{eq:P}
P(\mathbf{r}) = \dfrac{1}{2} \rho(\mathbf{r}) \langle |\mathbf{K}(\mathbf{r})|^2 \rangle,
\end{equation}
where angle brackets denote an ensemble averaging, which in this case is equivalent to a time-averaging due to the stationarity of the stochastic fields. The vector $\mathbf{K}(\mathbf{r})$ corresponds to the current density in $\mathbf{r}$.
The currents being a linear function of the incident fields, they can be decomposed according to the contribution of each incident plane wave:
\begin{equation}
\label{eq:K}
\mathbf{K}(\mathbf{r}) = \dfrac{1}{4\pi^2} \iint \underline{\underline{\tilde{K}}}^o(\mathbf{r}|\mathbf{k}_t) \cdot \mathbf{\tilde{E}}_i(\mathbf{k}_t)   d\mathbf{k}_t,
\end{equation}
with $\underline{\underline{\tilde{K}}}^o(\mathbf{r}|\mathbf{k}_t)$ a tensor describing the currents induced in $\mathbf{r}$ by incident plane waves of transverse wave-vector $\mathbf{k}_t$ and both possible polarizations. 
If one is interested in the total power absorbed by the structure, substituting (\ref{eq:K}) into (\ref{eq:P}) and integrating over the whole structure, we obtain:
\begin{equation}
\begin{split}
P = \dfrac{1}{32\pi^{4}} \iiint_\Omega& \rho(\mathbf{r}) \bigg\langle \Big( \iint \underline{\underline{\tilde{K}}}^o(\mathbf{r}|\mathbf{k}_t) \cdot \mathbf{\tilde{E}}_i(\mathbf{k}_t) d\mathbf{k}_t \Big)^{*} \\
&\cdot \Big(\iint \underline{\underline{\tilde{K}}}^o(\mathbf{r}|\mathbf{k}_t') \cdot \mathbf{\tilde{E}}_i(\mathbf{k}_t') d\mathbf{k}_t' \Big) \bigg\rangle d\mathbf{r},
\end{split}
\end{equation}
with $\iiint_\Omega$ the integration over the structure. Rearranging the  integration and averaging orders, and defining a double-dot operation according to
\begin{equation}
\mathbf{a}^{*} \cdot \underline{\underline{A}} \cdot \mathbf{b} = (\mathbf{a}^* \mathbf{b}) : \underline{\underline{A}},
\end{equation}
we obtain:
\begin{equation}
\label{eq:Ptot}
P = \dfrac{1}{16\pi^4} \iint \iint \Big\langle \mathbf{\tilde{E}}_i^*(\mathbf{k}_t) \mathbf{\tilde{E}}_i(\mathbf{k}_t') \Big\rangle 
: \underline{\underline{P_o}}(\mathbf{k}_t, \mathbf{k}_t') d\mathbf{k}_t d\mathbf{k}_t',
\end{equation}
with 
\begin{equation}
\label{eq:Po}
\underline{\underline{P_o}}(\mathbf{k}_t, \mathbf{k}_t') = \dfrac{1}{2}\iiint_\Omega \rho(\mathbf{r}) \underline{\underline{\tilde{K}}}^{o \dagger}(\mathbf{r}| \mathbf{k}_t) \cdot \underline{\underline{\tilde{K}}}^{o}(\mathbf{r}| \mathbf{k}_t') d\mathbf{r}
\end{equation}
the cross-spectral power density function, where $\underline{\underline{A}}^T$ and $\underline{\underline{A}}^{\dagger}$ are used to denote the transpose and Hermitian transpose of dyadic $\underline{\underline{A}}$ respectively. Equation (\ref{eq:Po}) is the extension of (6) in \citep{4}. The dyadic notation arises from the treatment of the polarization of the incident fields.

In Equation (\ref{eq:Ptot}), two different contributions can be highlighted: $\underline{\underline{P_o}}(\mathbf{k}_t, \mathbf{k}_t')$, which only depends on the structure,  and the ensemble average of the incident fields, which is independent of the structure. Both factors can be treated separately because the excitation is decoupled from the structure properties and geometry. It is worth noting that if we are only interested in the power absorbed in a given region of the structure (e.g. detectors), then the domain of integration of Equation (\ref{eq:Po}) can be limited to that region. 

\section{Coherent modes representation of the fields}
\label{sec:cohmodes}
In this section, we will focus on the partially coherent incident fields and expand them into a set of coherent modes. We start from $\underline{\underline{W}}(\mathbf{r}, \mathbf{r'})$, the cross spectral density dyadic \citep{1}, as defined by
\begin{equation}
\label{eq:w}
\underline{\underline{W}}(\mathbf{r}, \mathbf{r'}) = \langle \mathbf{E}_{i}^*(\mathbf{r}) \mathbf{E}_{i}(\mathbf{r'}) \rangle,
\end{equation}
which describes the state of spatial coherence of the incident fields.

It can be proven that the behaviour of the $\underline{\underline{W}}$ function is governed by the wave equations with respect to the coordinates $\mathbf{r}$ and $\mathbf{r'}$. Therefore, knowing the value of that function on a reference plane is sufficient to know the value of that function everywhere in space. However, propagating that function through an optical system or within the absorbing structure can be computationally intensive \citep{6}. An elegant way of easing the problem is to decompose the stochastic fields into a sum of non-stochastic (i.e. fully deterministic) modes, whose relative amplitudes are completely uncorrelated \citep{1,6}. Each of these modes can then be treated using classical coherent-field solvers. The corresponding expansion of $\underline{\underline{W}}$ reads 
\begin{equation}
\label{eq:mode_decomp}
\underline{\underline{W}}(\mathbf{r}, \mathbf{r'}) = \sum_p \lambda_p \mathbf{E}_p^*(\mathbf{r})  \mathbf{E}_p(\mathbf{r'}),
\end{equation}
with
\begin{equation}
\mathbf{E}_i(\mathbf{r}) = \sum_p a_p \mathbf{E}_p(\mathbf{r})
\end{equation}
\begin{equation}
\langle a^*_p a_q \rangle = \lambda_p \delta_{pq},
\end{equation}
where $\delta_{pq}$ is the Kronecker delta.
The propagation of the correlation function then corresponds to the independent propagation of each mode. 
The modes and their coefficients can be found solving the eigenvalue equation \citep{1}
\begin{equation}
\iint_S \underline{\underline{W}}(\mathbf{r}, \mathbf{r'}) \cdot \mathbf{E}_p^*(\mathbf{r'}) d\mathbf{r'} = \lambda_p \mathbf{E}_p^*(\mathbf{r}),
\end{equation}
This concept can be transferred to the spectral domain noticing that the relation between the spatial and the spectral domains is linear. Applying a Fourier transform to (\ref{eq:w}) and (\ref{eq:mode_decomp}) with respect to both $\mathbf{r}$ and $\mathbf{r}'$ coordinates, using the orthogonality of the Fourier kernel over the infinite plane and using definition (\ref{eq:18-09-01}), it can be shown that
\begin{equation}
\underline{\underline{\tilde{W}}}(\mathbf{k}_t, \mathbf{k}_t') = \sum_p \lambda_p \underline{\underline{\tilde{W}_p}}(\mathbf{k}_t, \mathbf{k}_t')
\end{equation}
\begin{equation}
\label{eq:15-01-03}
\underline{\underline{\tilde{W}_p}}(\mathbf{k}_t, \mathbf{k}_t') =
\mathbf{\tilde{E}}_p^*(\mathbf{k}_t)  \mathbf{\tilde{E}}_p(\mathbf{k}_t')
\end{equation}
with 
\begin{align}
\label{eq:wtilde}
&\underline{\underline{\tilde{W}}}(\mathbf{k}_t, \mathbf{k}_t') = \\
&	\begin{pmatrix}
	\langle \tilde{E}_{i,\text{TE}}^*(\mathbf{k}_t) \tilde{E}_{i,\text{TE}}(\mathbf{k}_t') \rangle
		& \langle \tilde{E}_{i,\text{TE}}^*(\mathbf{k}_t) \tilde{E}_{i,\text{TM}}(\mathbf{k}_t') \rangle \\
	\langle \tilde{E}_{i,\text{TM}}^*(\mathbf{k}_t) \tilde{E}_{i,\text{TE}}(\mathbf{k}_t') \rangle
		& \langle \tilde{E}_{i,\text{TM}}^*(\mathbf{k}_t) \tilde{E}_{i,\text{TM}}(\mathbf{k}_t') \rangle 
	\end{pmatrix} \nonumber.
\end{align}

Note that the phase factor of the Fourier transform of $E(\mathbf{r})$ with respect to $\mathbf{r}$ in both  (\ref{eq:w}) and (\ref{eq:mode_decomp}) is $\exp(-j \mathbf{k}_t \cdot \mathbf{r})$ instead of $\exp(j \mathbf{k}_t \cdot \mathbf{r})$ due to the complex conjugation.
This result can be introduced in Equation (\ref{eq:Ptot}) to yield:
\begin{equation}
\label{eq:Ptot_coh}
P = \dfrac{1}{16\pi^4} \sum_p \lambda_p \iint \iint \underline{\underline{\tilde{W}_p}}(\mathbf{k}_t, \mathbf{k}_t') 
: \underline{\underline{P_o}}(\mathbf{k}_t, \mathbf{k}_t') d\mathbf{k}_t d\mathbf{k}_t'.
\end{equation}

\section{Energy Absorption Interferometry}
Energy Absorption Interferometry (EAI) is an experimental technique used to characterize the way a structure can absorb electromagnetic fields \citep{3}. The Structure Under Test (SUT) is illuminated with two sources whose positions and relative phases can be controlled. Recording the powers absorbed for different positions and phase difference between sources, it is possible to predict the power absorbed by the structure when illuminated by any incident fields.

EAI can be implemented experimentally in many different ways, and over a wide range of wavelengths \citep{answer2}, depending on the way the SUT is illuminated (e.g: near-field probes, far field radiators) and the way the power absorbed by the sample is recorded. The power absorption can be measured directly when the SUT corresponds to photovoltaics and electromagnetic detectors, while, for other types of structures, the absorption can be recorded through the local temperature rise using bolometric calorimeters. A first demonstration has been reported at submillimeter-wavelengths \citep{answer1}.

\begin{figure}[!ht]
\center
\includegraphics[width = 8cm]{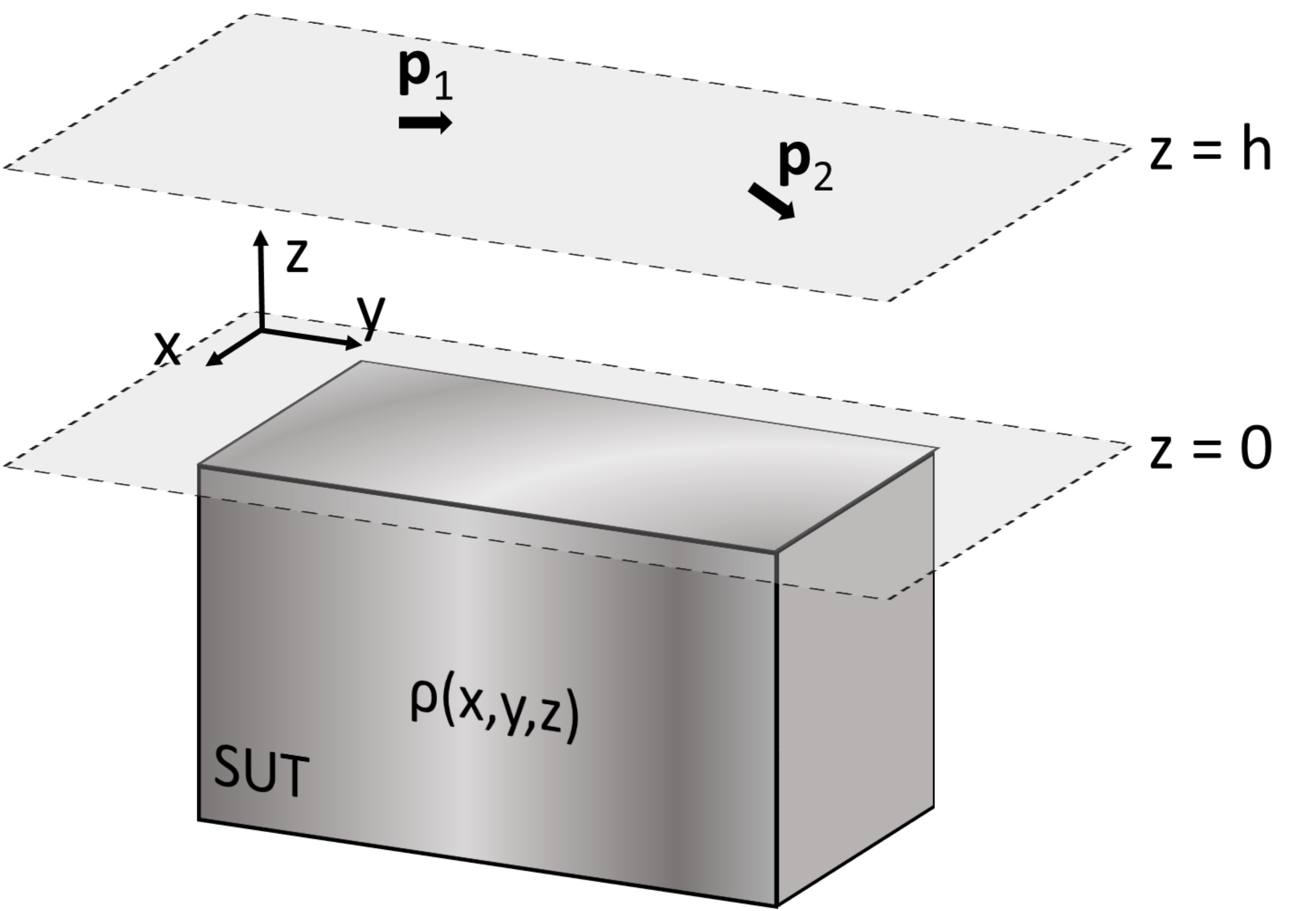}
\caption{Setup of the EAI. The two sources $\mathbf{p}_1$ and $\mathbf{p}_2$ are located in the $z=h$ plane. The power absorbed by the structure located in the half-space $z<0$ is recorded for each positions, orientations and relative phases of the two source.}
\label{fig:EAI}
\end{figure}

Consider two horizontal point sources corresponding to oscillating currents $\mathbf{J}_1(\mathbf{r})$ and $\mathbf{J}_2(\mathbf{r})$ such that
\begin{subequations}
\begin{align}
\mathbf{J}_1(\mathbf{r}) &= \mathbf{p}_1 \delta(\mathbf{r}_1) \\
\mathbf{J}_2(\mathbf{r}) &= \mathbf{p}_2 \delta(\mathbf{r}_2),
\end{align}
\end{subequations}
where $\delta$ is the Dirac delta function in 3D and $\mathbf{p}= (p_x; p_y; 0)$ where $p_i$ is the component of the current density in the direction $i$. The positions of the two sources, $\mathbf{r}_1$ and $\mathbf{r}_2$ respectively, both lie in the plane $z=h>0$. The relative phase between the sources can be controlled and their positions and orientations changed. The SUT is located in the half-space $z<0$. This experimental arrangement is illustrated in Figure \ref{fig:EAI}. 

We can define the matrix $\underline{\underline{S}}(\mathbf{k}_t|\mathbf{r})$ that provides the fields generated on the $z=0$ plane by a point source located at $\mathbf{r}$:
\begin{equation}
\label{eq:21-09-15-01}
\mathbf{\tilde{E}}_i(\mathbf{k}_t) \equiv \underline{\underline{S}}(\mathbf{k}_t|\mathbf{r}) \cdot \mathbf{p}.
\end{equation}
%
As a translation of the source in the spatial domain induces a linear phase slope in the angular spectrum domain, (\ref{eq:21-09-15-01}) simplifies to 
\begin{equation}
\label{eq:ptoE}
\mathbf{\tilde{E}}_i(\mathbf{k}_t) = \exp(j \mathbf{k}_t \cdot \mathbf{r}) \underline{\underline{S_o}}(\mathbf{k}_t) \cdot \mathbf{p}.
\end{equation}
$\underline{\underline{S_o}}(\mathbf{k}_t)$ is the plane-wave decomposition of the fields over $z=0$ produced by a source located at $\mathbf{r}=(0,0,h)$.
The different dyadic components are given by
\begin{equation}
S_{o,11} = \dfrac{\eta k_y k_0}{2 k_z k_t} \exp(-j k_z h)
\end{equation}
\begin{equation}
S_{o,12} = -\dfrac{\eta k_x k_0}{2 k_z k_t} \exp(-j k_z h)
\end{equation}
\begin{equation}
S_{o,21} = -\dfrac{\eta k_x}{2 k_t} \exp(-j k_z h)
\end{equation}
\begin{equation}
S_{o,22} = -\dfrac{\eta k_y}{2 k_t} \exp(-j k_z h),
\end{equation}
with $\eta = \sqrt{\mu_0/\varepsilon_0}$ the free space impedance. Substituting (\ref{eq:ptoE}) into (\ref{eq:K}), the currents induced in the structure by a point source $\mathbf{p}_1$ in $\mathbf{r}_1$ become
\begin{equation}
\label{eq:Krr_to_Krk}
\begin{split}
\mathbf{K}(\mathbf{r}) &\equiv ~ \underline{\underline{K}}^p(\mathbf{r}|\mathbf{r}_1) \cdot \mathbf{p}_1 \\
&=\dfrac{1}{4\pi^2} \iint \exp(j \mathbf{k}_t \cdot \mathbf{r}_1)
\underline{\underline{\tilde{K}^0}}(\mathbf{r}|\mathbf{k}_t) \cdot \underline{\underline{S_o}}(\mathbf{k}_t) \cdot \mathbf{p}_1 ~ d\mathbf{k}_t.
 \end{split}
\end{equation}

If the system is excited by two different point sources $\mathbf{p}_1$ and $\mathbf{p}_2$ located at positions $\mathbf{r}_1$ and $\mathbf{r}_2$ respectively, whose relative phase shift is $\phi$, starting from Equation (\ref{eq:P}) and using (\ref{eq:Krr_to_Krk}), the total power absorbed is
\begin{equation}
\begin{split}
P = \dfrac{1}{2}\iiint_\Omega \rho(\mathbf{r}) \Big( &\underline{\underline{K^p}}(\mathbf{r}|\mathbf{r}_1) \cdot \mathbf{p}_1 + \underline{\underline{K^p}}(\mathbf{r}|\mathbf{r}_2) \cdot \mathbf{p}_2 ~ e^{j\phi} \Big) \\
& \cdot \Big(\underline{\underline{K^p}}(\mathbf{r}|\mathbf{r}_1) \cdot \mathbf{p}_1 + \underline{\underline{K^p}}(\mathbf{r}|\mathbf{r}_2) \cdot \mathbf{p}_2 ~ e^{j\phi} \Big) d\mathbf{r}.
\end{split}
\end{equation}

Distributing the product, separating the integrals and using the same compact notation as in Equation (\ref{eq:Ptot}), it is possible to extend Equation (16) of \citep{4} to 3D geometries and polarized excitations:
\begin{equation}
P= \dfrac{1}{2} (\alpha_{11} + \alpha_{22} + 2\alpha_{12} \cos(\phi-\beta_{12})),
\end{equation}
with
\begin{equation}
\alpha_{pq} ~ e^{j\beta_{pq}} = \iiint_\Omega \Big( \rho(\mathbf{r}) \underline{\underline{K}}^{p \dagger}(\mathbf{r}|\mathbf{r}_p) \cdot \underline{\underline{K}}^{p}(\mathbf{r}|\mathbf{r}_q) \Big) d\mathbf{r}
 : \Big(\mathbf{p}^*_p \mathbf{p}_q \Big)
\end{equation}
and $\beta_{pp} = 0$. The correlation function $C_{pq}(y_1, y_2)$ defined in \citep{4} can then be extended to 3D according to
\begin{equation}
\label{eq:C}
\underline{\underline{C}}(\mathbf{r}_1, \mathbf{r}_2) = \iiint_\Omega \Big( \rho(\mathbf{r}) \underline{\underline{K}}^{p \dagger}(\mathbf{r}|\mathbf{r}_1) \cdot \underline{\underline{K}}^{p}(\mathbf{r}|\mathbf{r}_2) \Big) d\mathbf{r}.
\end{equation}
Using EAI experiments, the magnitude and phase of $\underline{\underline{C}}(\mathbf{r}_1, \mathbf{r}_2)$ can be obtained varying the relative phase of the sources and measuring the oscillating pattern in the absorption.

\section{Relation between EAI and spectral correlation}
In this section, we will provide a link between the data that can be gathered using EAI, which corresponds to the spatial correlation function $\underline{\underline{C}}(\mathbf{r}_1, \mathbf{r}_2)$, and the cross-spectral power density function $\underline{\underline{P_o}}(\mathbf{k}_t, \mathbf{k}_t')$ as defined in (\ref{eq:Po}). Starting from the definition of the correlation function given in (\ref{eq:C}), we can use (\ref{eq:Krr_to_Krk}) to obtain
\begin{equation}
\label{eq:c12cumb}
\begin{split}
\underline{\underline{C}}(\mathbf{r}_1, \mathbf{r}_2) = \dfrac{1}{(2\pi)^4} \iint \iint \exp(-j \mathbf{k}_t \cdot \mathbf{r}_1) \underline{\underline{S_o}}^{\dagger}(\mathbf{k}_t)  \\
\cdot \iiint_\Omega \rho(\mathbf{r}) ~ \underline{\underline{\tilde{K}}}^{o \dagger}(\mathbf{r}|\mathbf{k}_t) \cdot \underline{\underline{\tilde{K}}}^{o}(\mathbf{r}|\mathbf{k}_t') d\mathbf{r} \\
 \cdot \underline{\underline{S_o}}(\mathbf{k}_t') \exp(j \mathbf{k}_t' \cdot \mathbf{r}_2) d\mathbf{k}_t' d\mathbf{k}_t.
\end{split}
\end{equation}
We simplify this expression using (\ref{eq:Po}) and the change of variables
\begin{equation}
\label{eq:r+-}
\begin{split}
\mathbf{r}^+ = \dfrac{\mathbf{r}_2 + \mathbf{r}_1}{2} ~~~~~~& \mathbf{r}^- = \mathbf{r}_1 - \mathbf{r}_2 \\
\mathbf{k}_t^+ = \dfrac{\mathbf{k}_t + \mathbf{k}_t'}{2} ~~~~~~& \mathbf{k}_t^- = \mathbf{k}_t - \mathbf{k}_t',
\end{split}
\end{equation}
to yield
\begin{equation}
\label{eq:Cpresquetot}
\begin{split}
\underline{\underline{C}}(\mathbf{r}_1, \mathbf{r}_2) = \dfrac{1}{8\pi^4} &\iint \iint \underline{\underline{S_o}}^\dagger(\mathbf{k}_t) \cdot  \underline{\underline{P_o}}(\mathbf{k}_t, \mathbf{k}_t') \cdot \underline{\underline{S_o}}(\mathbf{k}_t') \\
\times& \exp(-j\mathbf{k}_t^- \cdot  \mathbf{r}^+) \exp(-j \mathbf{k}_t^+ \cdot \mathbf{r}^-) d\mathbf{k}_t^+ d\mathbf{k}_t^-.
\end{split}
\end{equation}
As in \citep{4}, we can notice that Equation (\ref{eq:Cpresquetot}) corresponds to a double inverse Fourier transform. Therefore, using the relation between a function and its Fourier transform, we finally obtain
\begin{equation}
\label{eq:PoC12}
\begin{split}
\underline{\underline{P_o}}(\mathbf{k}_t, \mathbf{k}_t') = \dfrac{1}{2} \big(\underline{\underline{S_o}}^{-1}&(\mathbf{k}_t)\big)^{\dagger} \\
\cdot \iint \iint \underline{\underline{C}}^{\Delta}(\mathbf{r}^+, \mathbf{r}^-) & ~ e^{j (\mathbf{k}_t^- \cdot \mathbf{r}^+ + \mathbf{k}_t^+ \cdot \mathbf{r}^-)} d\mathbf{r}^+ d\mathbf{r}^- \cdot \underline{\underline{S_o}}^{-1}(\mathbf{k}_t'),
\end{split}
\end{equation}
with
\begin{equation}
\underline{\underline{C}}^\Delta(\mathbf{r}^+, \mathbf{r}^-) = \underline{\underline{C}}\Big(\mathbf{r}^+ +\dfrac{\mathbf{r}^-}{2}, \mathbf{r}^+ -\dfrac{\mathbf{r}^-}{2} \Big).
\end{equation}
To summarize the results, if the spatial spectrum of the sources is known (through the matrix $\underline{\underline{S_o}}$), one has to apply a Fourier transform on the measurements to obtain the cross correlation function in the spectral domain.

\section{Dissipation in periodic structures}
\label{sec:6}
The computation of the total power absorbed by a structure using formula (\ref{eq:Ptot}) requires the evaluation of several integrals. The first one is a volume integral over the structure. The two others are 2D integrals over the plane wave spectrum of the incident light. In case the structure studied is periodic, these expressions can be simplified. Only the case of 2D periodicity of period $a$ and $b$ along the $\hat{x}$ and $\hat{y}$ directions will be considered hereafter. The extension to any other type of periodicity is straightforward.

Let us qualify as \textit{quasiperiodic} an excitation which shares the same periodicity as the structure within a linear phase shift.
Using the Floquet theorem, it can be shown that the currents $\underline{\underline{K}}^{o}(\mathbf{r}| \mathbf{k}_t)$ induced on a periodic structure by a quasiperiodic excitation share the same quasiperiodicity as the excitation. This means that the product $\rho(\mathbf{r}) \underline{\underline{\tilde{K}}}^{o \dagger}(\mathbf{r}| \mathbf{k}_t) \cdot \underline{\underline{\tilde{K}}}^{o}(\mathbf{r}| \mathbf{k}_t') \exp\big(j(\mathbf{k}_t'-\mathbf{k}_t)\big)$, which appears in Equation (\ref{eq:Po}), is periodic with periods $a$ and $b$ in the $\mathbf{\hat{x}}$ and $\mathbf{\hat{y}}$ directions, respectively, and can be decomposed into a Fourier series
\begin{equation}
\begin{split}
\underline{\underline{P_o}}(\mathbf{k}_t, \mathbf{k}_t') = \dfrac{1}{2}\iiint_\Omega &\exp\big(-j(\mathbf{k}_t'-\mathbf{k}_t) \cdot \mathbf{r})\\
\times \sum_{m,n =-\infty}^{\infty} \underline{\underline{Q_{mn}}}(z|&\mathbf{k}_t, \mathbf{k}_t') \exp(-j 2\mathbf{k}_t^{mn} \cdot \mathbf{r}) d\mathbf{r}
\end{split}
\end{equation}
with 
\begin{equation}
\begin{split}
 \underline{\underline{Q_{mn}}}(z|\mathbf{k}_t, \mathbf{k}_t') = \dfrac{1}{ab}\iint_{0,0}^{a,b} 
  \rho(\mathbf{r}) ~ \underline{\underline{\tilde{K}}}^{o \dagger}(\mathbf{r}| \mathbf{k}_t) \cdot \underline{\underline{\tilde{K}}}^{o}(\mathbf{r}| \mathbf{k}_t')\\
\times\exp\big(j(\mathbf{k}_t'-\mathbf{k}_t) \cdot \mathbf{r})
\exp(j 2\mathbf{k}_t^{mn} \cdot \mathbf{r}) dy ~ dx,
\end{split}
 \end{equation}
\begin{equation}
\mathbf{k}_t^{mn} = 
\begin{pmatrix}
m \dfrac{\pi}{a}, &
n \dfrac{\pi}{b}, &
0
\end{pmatrix} ~~~~~~ m,n \in \mathbb{Z}.
\end{equation}

These expressions can be used in Equation (\ref{eq:Ptot_coh}). Changing the order of summation and integration and noticing that $\iint \exp(j\mathbf{k}) d\mathbf{k} = (2\pi)^2 \delta(\mathbf{k})$, the mean power dissipated in each unit cell (\ref{eq:Ptot_coh}) can be expressed as
\begin{equation}
\label{eq:16-09-15-1}
\begin{split}
P = \dfrac{1}{4\pi^2} \sum_p \lambda_p \sum_{m,n} \iint \underline{\underline{\tilde{W}_p}}(\mathbf{k}_t - \mathbf{k}_{t}^{mn}, \mathbf{k}_t+\mathbf{k}_{t}^{mn}) \\
: \underline{\underline{\mathcal{H}_{mn}}}(\mathbf{k}_t) d\mathbf{k}_t ,
\end{split}
\end{equation}
\begin{equation}
\begin{split}
\label{eq:17-09-01}
\underline{\underline{\mathcal{H}_{mn}}}(\mathbf{k}_t) &= \dfrac{1}{2} \\ \times \iiint_{\Omega_0}& \rho(\mathbf{r}) \underline{\underline{\tilde{K}}}^{o \dagger}(\mathbf{r}| \mathbf{k}_t- \mathbf{k}_{t}^{mn}) \cdot \underline{\underline{\tilde{K}}}^{o}(\mathbf{r}| \mathbf{k}_t+\mathbf{k}_{t}^{mn}) d\mathbf{r},
\end{split}
\end{equation}
with $\Omega_0$ the intersection of the volume $\Omega$ with the volume of one unit cell of the structure.

As mentioned in \citep{4} for the 2D case, both $\underline{\underline{\tilde{W}_p}}(\mathbf{k}_t-\mathbf{k}_{t}^{mn}, \mathbf{k}_t+\mathbf{k}_{t}^{mn})$ and $\underline{\underline{\mathcal{H}_{mn}}}(\mathbf{k}_t)$ tend to vanish for large spatial frequencies or large $m,n$ indices. Therefore the summation and the spectral integration can be truncated. The upper values to consider depend on the distance that the fields have to travel before reaching the structure, and the depth of penetration of the fields inside the structure. Thus, they depend on the geometry of the structure studied.

$\mathbf{k}_{t}^{mn}$ takes discrete values, so $\underline{\underline{P_o}}(\mathbf{k}_t, \mathbf{k}_t')$ is also discrete with respect to the difference of spatial frequencies. As underlined in Equation (\ref{eq:PoC12}), a Fourier relation links $\underline{\underline{P_o}}$ to the data collected by the EAI, which is invariant w.r.t. a shift by an integer number of cells. Therefore, $\underline{\underline{C}}^{\Delta}(\mathbf{r}^+, \mathbf{r}^-)$ is periodic with respect to $\mathbf{r}^+$. 
Consequently, to fully characterize a periodic structure, one of the two sources only has to scan one unit-cell, while the other one still has to scan the whole structure, decreasing the total amount of data needed.

A final observation concerns the computation of the power absorbed by an infinite periodic structure excited by a non-periodic source. The non-periodic source can always be decomposed into a sum (or integral) of quasiperiodic sources \citep{asm_capo}, each quasiperiodic component being periodic within a different phase factor. Then, the total power absorbed by the structure corresponds to the sum of the powers absorbed by the structure for each quasiperiodic component of the source taken separately. Indeed, the interference between different quasiperiodic components of the source will vanish when summing the power dissipated by each unit cell of the infinite structure. However, if one is interested in the power dissipated within one unit cell, the local interference between the different quasiperiodic components of the source does not vanish anymore. Thus, equation (\ref{eq:Ptot_coh}) must be used instead of Equation (\ref{eq:16-09-15-1}). Adapting the phase of the correlation between the different quasiperiodic components of the source, it is possible to compute the power dissipated in any unit-cell (see (\ref{eq:Po})).

\section{Evolution of \texorpdfstring{$H_{mn}$}{Hmn} with height}

In most practical cases, the source of the incident fields (e.g: thermal source or optical system) is not touching the absorbing structure. The incident fields then have to propagate to reach the structure. Consider an incident field $\mathbf{\tilde{E}}_i(h|\mathbf{k}_t)$ in $z=h>0$. In $z=h'<h$, the fields will change according to the factor
\begin{equation}
\mathbf{\tilde{E}}_i(h'|\mathbf{k}_t) = \mathbf{\tilde{E}}_i(h|\mathbf{k}_t) \exp\big( -j k_z (h-h') \big).
\end{equation}
Using Equation (\ref{eq:Ptot}), we then obtain the dissipated power as 
\begin{equation}
\begin{split}
P = \dfrac{1}{32\pi^4} \iint \iint \Big\langle \mathbf{\tilde{E}}_i^*(&\mathbf{k}_t) \mathbf{\tilde{E}}_i(\mathbf{k}_t') \Big\rangle \\
\times \exp\big(-j (k_z&+k_z') h\big): \underline{\underline{P_o}}(\mathbf{k}_t, \mathbf{k}_t') d\mathbf{k}_t d\mathbf{k}_t',
\end{split}
\end{equation}
where
\begin{equation}
k'_z = \pm \sqrt{k_0^2 - {k}_t'^2},
\end{equation}
the sign of $k'_z$ being subject to the same convention as previously defined.
We can then introduce the notation $\underline{\underline{P}}(z|\mathbf{k}_t, \mathbf{k}_t')$, which corresponds to the cross-correlation function for absorption if the fields are emitted in $z>0$
\begin{equation}
\label{eq:24-07_02}
\underline{\underline{P}}(z|\mathbf{k}_t, \mathbf{k}_t') = \exp\big(-j (k_z+k_z') z\big) ~ \underline{\underline{P_o}}(\mathbf{k}_t, \mathbf{k}_t').
\end{equation}
Similarly, for the periodic case, we can introduce 
\begin{equation}
\label{eq:24-07_01}
\begin{split}
\underline{\underline{\mathcal{H}_{mn}}}(z|\mathbf{k}_t+\mathbf{k}_{t}^{mn}) = 
&\exp\big( -j (k_z+k_z^{mn}) z \big) ~ \underline{\underline{\mathcal{H}_{mn}}}(\mathbf{k}_t+\mathbf{k}_{t}^{mn})
\end{split}
\end{equation}
with
\begin{equation}
k_z^{mn} = \pm \sqrt{k_0^2-|\mathbf{k}_t+2\mathbf{k}_t^{mn}|^2}.
\end{equation}
The sign of $k_z^{mn}$ following the conventions already defined for $k_z$ in (\ref{eq:08-07-01}).
The main consequence of Equations (\ref{eq:24-07_02}) and (\ref{eq:24-07_01}) is that the amplitude of the correlation function decreases exponentially with respect to the distance between the sources and the structure in the evanescent part of the spectrum. It means that, for a predefined accuracy, the farther the sources, the smaller the domain on which the $\underline{\underline{P}}$ and $\underline{\underline{\mathcal{H}_{mn}}}$ functions need to be actually tabulated. 

Moreover, if we consider that $k_t \gg k_0$, we have 
\begin{equation}
k_z = -\sqrt{k_0^2-k_t^2} \simeq -\sqrt{-k_t^2} = -j k_t.
\end{equation}
Therefore, for high $(m,n)$ indices, we have 
\begin{equation}
\begin{split}
\Big|&
\underline{\underline{\mathcal{H}_{mn}}}(z|\mathbf{k}_t+\mathbf{k}_{t}^{mn})
\Big| \\
 &=
\Big| \exp(-j(k_z+k_z^{mn}) z ) \underline{\underline{\mathcal{H}_{mn}}}(\mathbf{k}_t+\mathbf{k}_{t}^{mn})
\Big| \\
&\simeq
\Big| \exp(-j k_z z) \exp(- |2\mathbf{k}_t^{mn}+\mathbf{k}_t| z)  \underline{\underline{\mathcal{H}_{mn}}}(\mathbf{k}_t+\mathbf{k}_{t}^{mn})
\Big| \\
& \leq
\Big| \exp\big(- (2k_t^{mn}-2k_0) z\big)  \underline{\underline{\mathcal{H}_{mn}}}(\mathbf{k}_t+\mathbf{k}_{t}^{mn})
\Big|
\end{split}
\end{equation}

This means that once a certain distance between the sources and the structure is exceeded, the high order correlation functions then decrease exponentially with respect to $\sqrt{m^2+n^2}$. 
Consequently, the further the sources are from the absorbing device, the smaller the number of correlations functions required to characterize the absorber's behaviour, and the smaller the domain over which these functions must be tabulated.  Since the results from EAI and the correlation function are related by a Fourier transform, it also means that the further the sources are from the structure, the coarser the spatial sampling of the data can be.

\section{Computation of the absorption modes}
\label{sec:8}
As suggested in \citep{3} and implemented in \citep{absorptionmodes} for a 2D geometry, this mathematical framework can be used to compute the modes through which a structure can absorb power. The total power dissipated by the structure corresponds to the sum of the powers dissipated by each of these modes separately.

Equation (\ref{eq:Ptot}) can be restated within the theory of the linear operators of Hilbert spaces. We consider the functional space made of all the possible incident field distributions and define the scalar product between two fields distributions $\mathbf{\tilde{E}}_p(\mathbf{k}_t)$ and $\mathbf{\tilde{E}}_q(\mathbf{k}_t)$ as follows:
\begin{equation}
\big[ \mathbf{E}_p | \mathbf{E}_q \big] = \iint \mathbf{E}_p^\dagger(\mathbf{k}_t) \cdot \mathbf{E}_q(\mathbf{k}_t) ~ d\mathbf{k}_t.
\end{equation}
An unconventional `bra-ket' notation has been used to avoid
confusion with $\langle \mathbf{E}_p  \mathbf{E}_q \rangle$, which corresponds to the ensemble average of the fields. For any physical fields distribution, this scalar product is sesquilinear, positive-definite and exhibits a conjugate symmetry (i.e. $\big[ \mathbf{\tilde{E}}_p | \mathbf{\tilde{E}}_q \big] = \big[ \mathbf{\tilde{E}}_q | \mathbf{\tilde{E}}_p \big]^*$). In the framework of this Hilbert space, we can define the operator $\hat{P}$ such that
\begin{equation}
|\hat{P} \mathbf{\tilde{E}} \big] = \dfrac{1}{16\pi^4} \iint \underline{\underline{P_o}}(\mathbf{k}_t, \mathbf{k}_t') \cdot \mathbf{\tilde{E}}(\mathbf{k}_t') ~ d\mathbf{k}_t'.
\end{equation}
Equation (\ref{eq:Ptot}) then reads
\begin{equation}
\label{eq:15-01-02}
P = \big\langle \big[ \mathbf{\tilde{E}}_i | \hat{P} \mathbf{\tilde{E}}_i \big] \big\rangle.
\end{equation}
Using the definition of $\underline{\underline{P_o}}$ in (\ref{eq:Po}), it can be shown that the operator $\hat{P}$ is Hermitian, so that its eigenvalues $\Lambda_\alpha$ are real and the corresponding eigenvectors $\mathcal{E}_\alpha(\mathbf{k}_t)$ are orthogonal. From a physical point of view, one of the main consequences is that the power dissipated by any incident field $\mathbf{\tilde{E}}_i(\mathbf{k}_t)$ is 
\begin{equation}
\label{eq:18-12-01}
P =  \sum_\alpha \Lambda_\alpha \Big| \big[ \mathcal{E}_\alpha| \mathbf{\tilde{E}}_i \big] \Big|^2,
\end{equation}
with $|a|$ the magnitude of $a$. Therefore, the power is always real-valued, as required by physical law. Moreover, from a mathematical point of view, it means that a simple orthogonal projection of the incident fields on the different modes is required to obtain the total power absorbed. Therefore, once all the modes with a non-negligible eigenvalue are known, the exact projection of the incident fields into these modes does not require any knowledge about the other modes. This is convenient in the sense that only the incident fields are required, and not the total fields (incident and reflected).
Last, for finite structures, it can be shown that the $\hat{P}$ operator is Hilbert-Schmidt and therefore compact. It means that the number of modes whose contribution to absorption is not negligible is finite. Moreover, in the case of structures that are small with respect to the wavelength of operation, this number is generally small.

From a numerical point of view, the $\hat{P}$ operator has to be discretized to compute these modes through an eigenvector decomposition. It can also be proven that, in the case of a periodic structure, each absorption mode is only composed of Floquet modes characterized by a common phase shift between consecutive unit cells. Therefore, the absorption modes of a periodic structure can be computed by simulating just one unit cell with a periodic solver and the appropriate periodic boundary conditions. This results in a naturally discretized $\hat{P}$ operator, such that the numerical solution of the problem is straightforward.

As mentioned previously, only the modes whose contribution to absorption is non-negligible need to be kept, resulting in a significant reduction in the amount of data that must be stored. The data required to fully characterize the absorption by non-periodic structures reduces from a 4D function to a compact set of 2D functions corresponding to the absorption modes $\mathcal{E}_\alpha(\mathbf{k}_t)$ for $\alpha = 1, 2, ...$ and $\mathbf{k}_t \in \mathcal{R}^2$. For periodic structures, a discrete set of discrete functions has to be stored for each possible phase shift between consecutive unit cells. That data can be reorganized into a discrete set of 2D functions, where each function is defined over the first Brillouin zone. Generally, both sets for periodic and non-periodic structures are small if the structure or its unit cell are not large with respect to the wavelength of the incident fields.

\section{Absorption by a structure}

Now that all necessary operators are available, we can address the problem of power absorption by a periodic structure illuminated by a spatially partially coherent field. However, we will first briefly summarize the results obtained so far.

In Section \ref{sec:cohmodes}, we considered the treatment of spatially partially coherent fields by decomposition into coherent modes.  The knowledge of the response of the structure to coherent fields is then sufficient to calculate the absorbed power.  In Section \ref{sec:2}, we provided a alternative scheme for characterizing power absorption by any structure, based on the computation of the 4D cross-spectral power density function $\underline{\underline{P_o}}(\mathbf{k}_t, \mathbf{k}_t')$. Then, in Section \ref{sec:6}, we showed that, provided that the structure is periodic, a 2D compact set of 2D functions $\underline{\underline{\mathcal{H}_{mn}}}(\mathbf{k}_t)$ is sufficient to fully characterize the way in which the periodic structure can absorb fields.

Once this information has been retrieved using simulations or experiments, the collected data needs to be stored. The data can be reduced in size by the use of the so-called absorption modes (cf. Section \ref{sec:8}).

Expanding Equation (\ref{eq:18-12-01}), the total absorbed power for any coherent excitation becomes
\begin{equation}
\begin{split}
P = \sum_\alpha \Lambda_\alpha \iint &\mathcal{E}_\alpha^\dagger(\mathbf{k}_t) \cdot \mathbf{\tilde{E}}_i(\mathbf{k}_t) d\mathbf{k}_t   \iint \mathbf{\tilde{E}}_i^\dagger(\mathbf{k}_t') \cdot \mathcal{E}_\alpha(\mathbf{k}_t') d\mathbf{k}_t'.
\end{split}
\end{equation}
Rearranging this expression and noticing that a partially coherent excitation can be rephrased as a sum of coherent excitations that are orthogonal to each other, it can be shown that the total power absorbed by a structure when it is illuminated by partially coherent fields is
\begin{equation}
\label{eq:15-01-04}
P = \sum_\alpha \Lambda_\alpha \iint \iint \mathcal{E}_\alpha^\dagger(\mathbf{k}_t) \cdot \underline{\underline{\tilde{W}}}(\mathbf{k}_t, \mathbf{k}_t') \cdot \mathcal{E}_\alpha(\mathbf{k}_t') d\mathbf{k}_t d\mathbf{k}_t'.
\end{equation}
In compact notation, this becomes 
\begin{equation}
\label{eq:15-01-01}
P = \sum_\alpha \Lambda_\alpha \big[ \mathcal{E}_\alpha | \hat{W} \mathcal{E}_\alpha \big].
\end{equation}

Notice the resemblance between (\ref{eq:15-01-01}) and (\ref{eq:15-01-02}). Modes $\mathcal{E}_\alpha$ and $\mathbf{\tilde{E}}_p$ are obtained using an eigenvalue decomposition of the operators $\hat{P}$ and $\hat{W}$, respectively. One can see that, provided that this eigenvalue decomposition has been performed for at least one of the two operators, the partially coherent nature of the fields can be handled using either (\ref{eq:Ptot_coh}) or (\ref{eq:15-01-04}). If one wants to study the way in which given partially coherent fields can be absorbed by any structure, the eigenvalue decomposition of the incident fields, which is independent from the structure, can be performed once and the power absorbed by each structure computed using (\ref{eq:Ptot_coh}), while if one is interested in the way a given structure can absorb any partially coherent incident fields, it is more interesting to compute the absorption modes of the structure, which are independent from the incident fields, and apply (\ref{eq:15-01-04}) to compute the power absorbed for each excitation.

\section{Numerical results}
\begin{figure}[!ht]
\center
\includegraphics[width = 8cm]{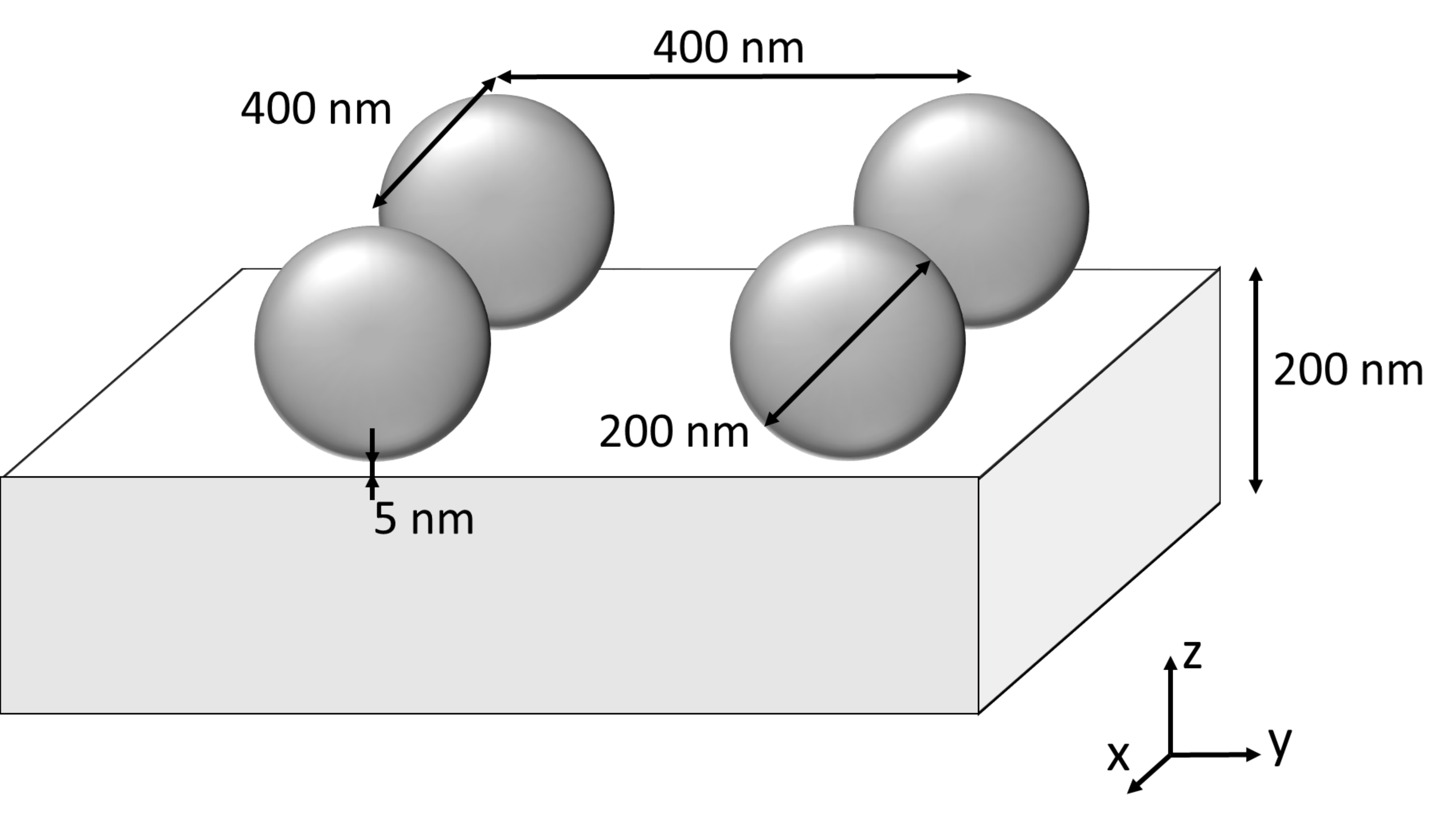}
\caption{Structure used for the simulations: an infinite array of golden nanospheres deposited on a suspended membrane.}
\label{fig:struct_simu}
\end{figure}

To illustrate the technique, we computed the $\underline{\underline{\mathcal{H}_{mn}}}$ functions for a periodic structure mimicking the typical design of room-temperature bolometers \citep{bolometers}. The structure consists of a squared lattice of gold nanospheres used as an absorbing layer deposited on a suspended membrane. The membrane is 200 nm thick and the radius of the spheres is 100 nm. The spacing between two consecutive spheres is 200 nm. 
We positioned the spheres 5 nm above the surface of the layer to avoid some numerical issues. However, this distance being smaller that $1/100^\text{th}$ of the wavelength, it has a negligible impact on the final results.
The free-space wavelength of the incident fields is 750 nm, which corresponds to a maximum of absorption for a 100 nm free-standing single golden sphere \citep{simu1}. Plasmonic effects of gold at that frequency are taken into account by using a negative value for the real part of the permittivity, while losses mechanisms are taken into account adding a negative imaginary part to the permittivity. At this wavelength, the complex relative permittivity of gold is then $\varepsilon_r = -21.2 -j1.56$ \citep{simu2} and the permittivity of the suspended membrane is $\varepsilon_r = 4$. The structure can be seen on Figure \ref{fig:struct_simu}. The $z=0$ plane coincides with the top of the spheres.

The coherent solver used to perform the full-wave simulation of the structure is based on the Poggio-Miller-Chan-Harrington-Wu-Tsay (PMCHWT) formulation applied to the Method of Moments \citep{mom1, mom2, mom3}. The periodicity of the structure is handled through the use of the periodic Green's function \citep{mom4}. The implementation of the code is described in \citep{mom5}. The sphere and the interfaces between the layer and the air are discretized using 690 Rao-Wilton-Glisson (RWG) \citep{RWG} and 800 rooftop basis functions, respectively.

\begin{figure*}
\center
\begin{tabular}{cc}
Power absorption for TE plane waves (norm) & Power absorption for TM plane waves (norm)\\
\includegraphics[height = 5.5cm]{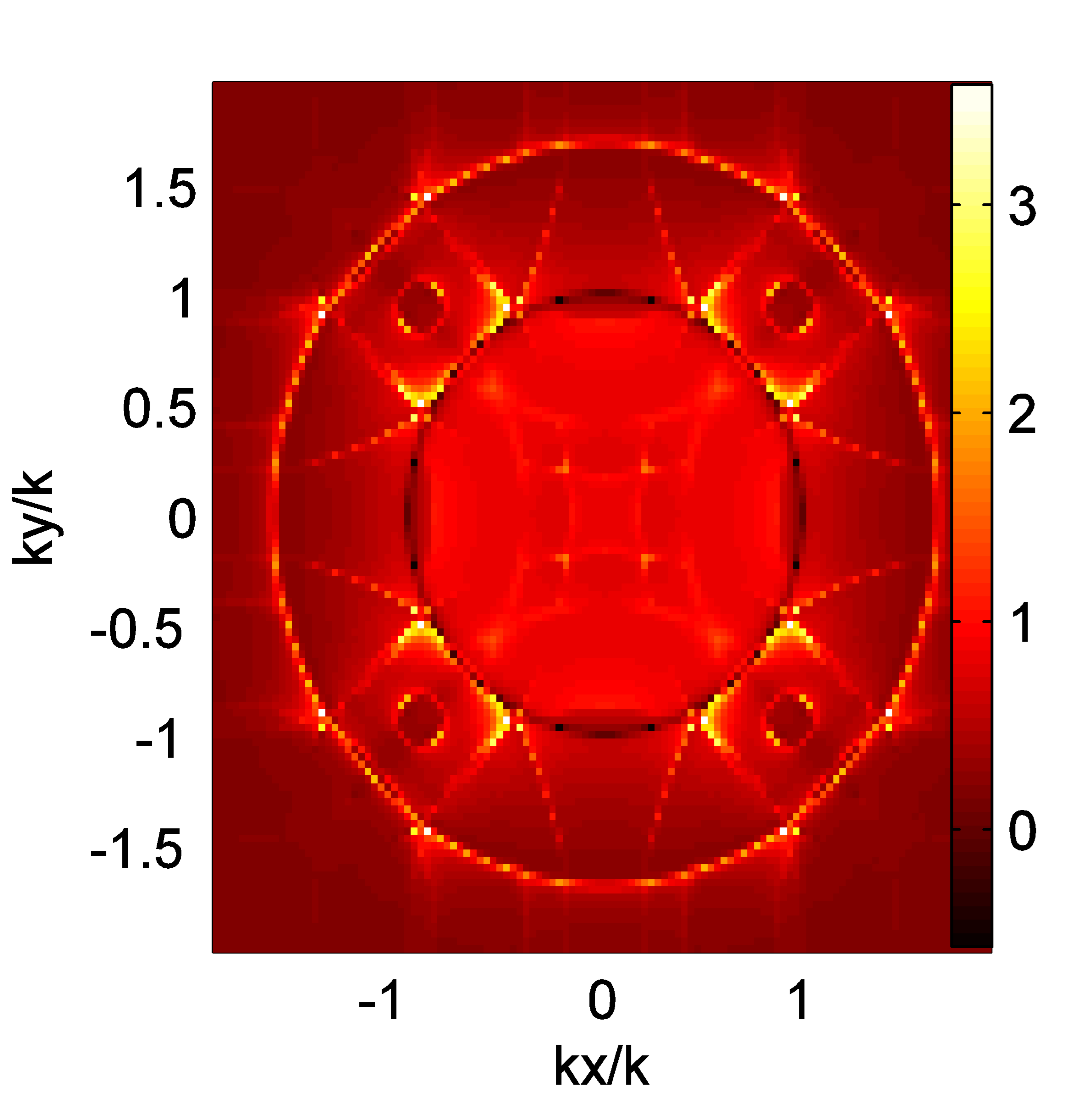} &
\includegraphics[height = 5.5cm]{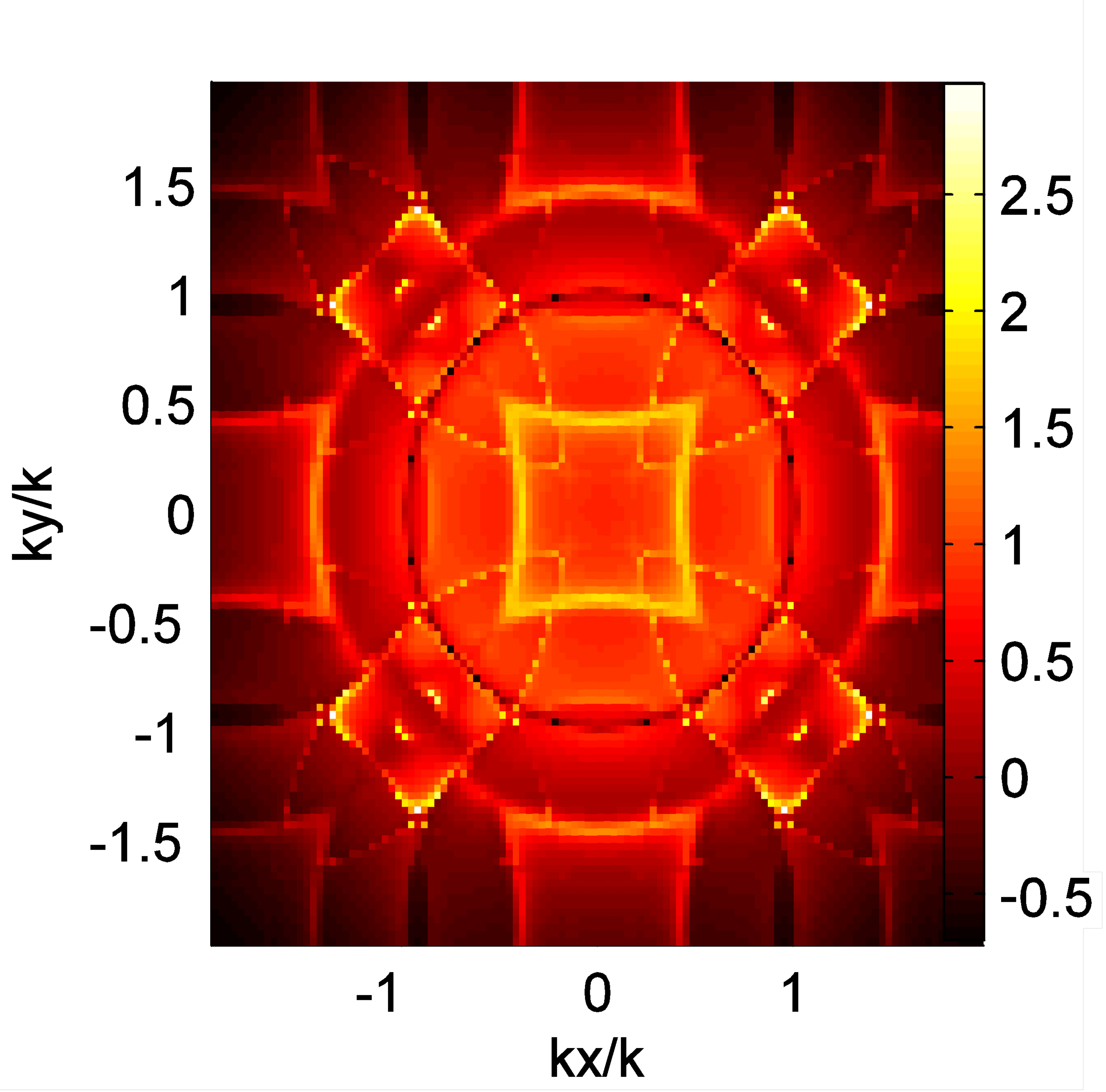} \\
~~~~~~~(a) & ~~~(b) \\
~~ \\
\multicolumn{2}{c}{Coupling between TE and TM plane waves of identical wave-vectors (norm and phase)} \\
\includegraphics[height = 5.5cm]{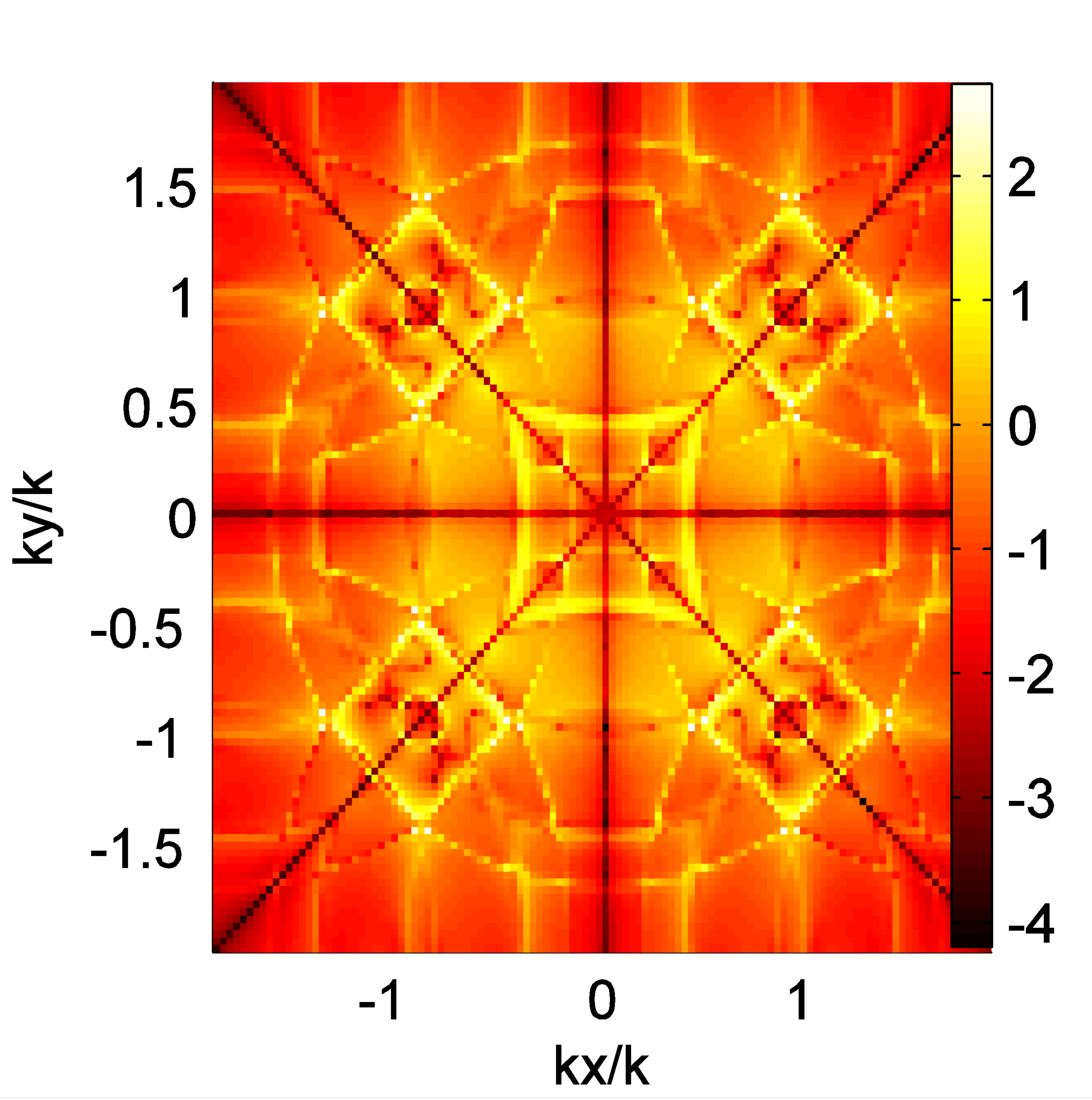} &
\includegraphics[height = 5.5cm]{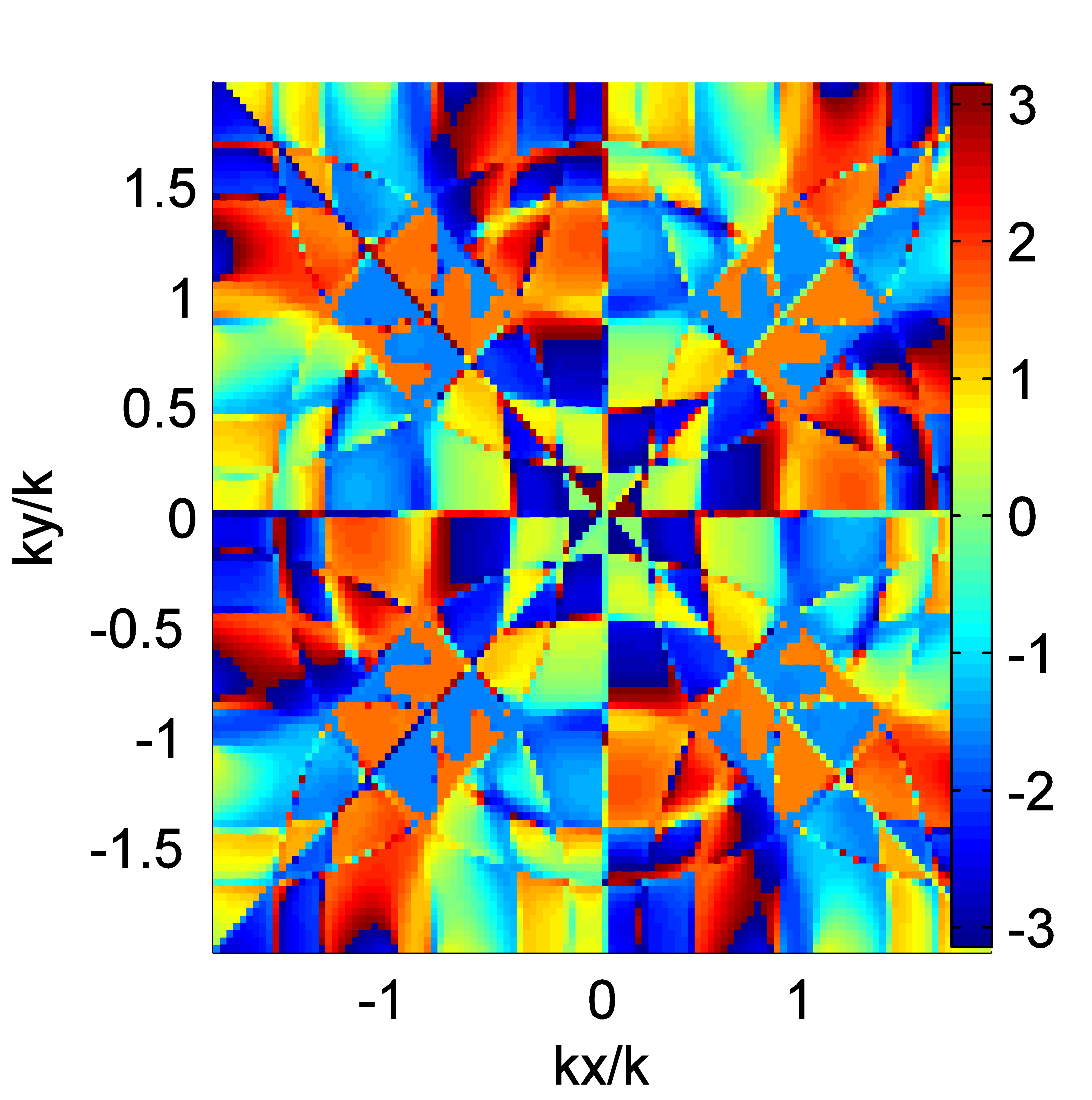} \\
~~~~~~(c) & ~~~~~~(d) \\
~~\\
\multicolumn{2}{c}{Coupling between TE waves of wave vectors $\mathbf{k}_t-\mathbf{k}_{t,01}$ and $\mathbf{k}_t+\mathbf{k}_{t,01}$ (norm and phase)} \\
\includegraphics[height = 5.5cm]{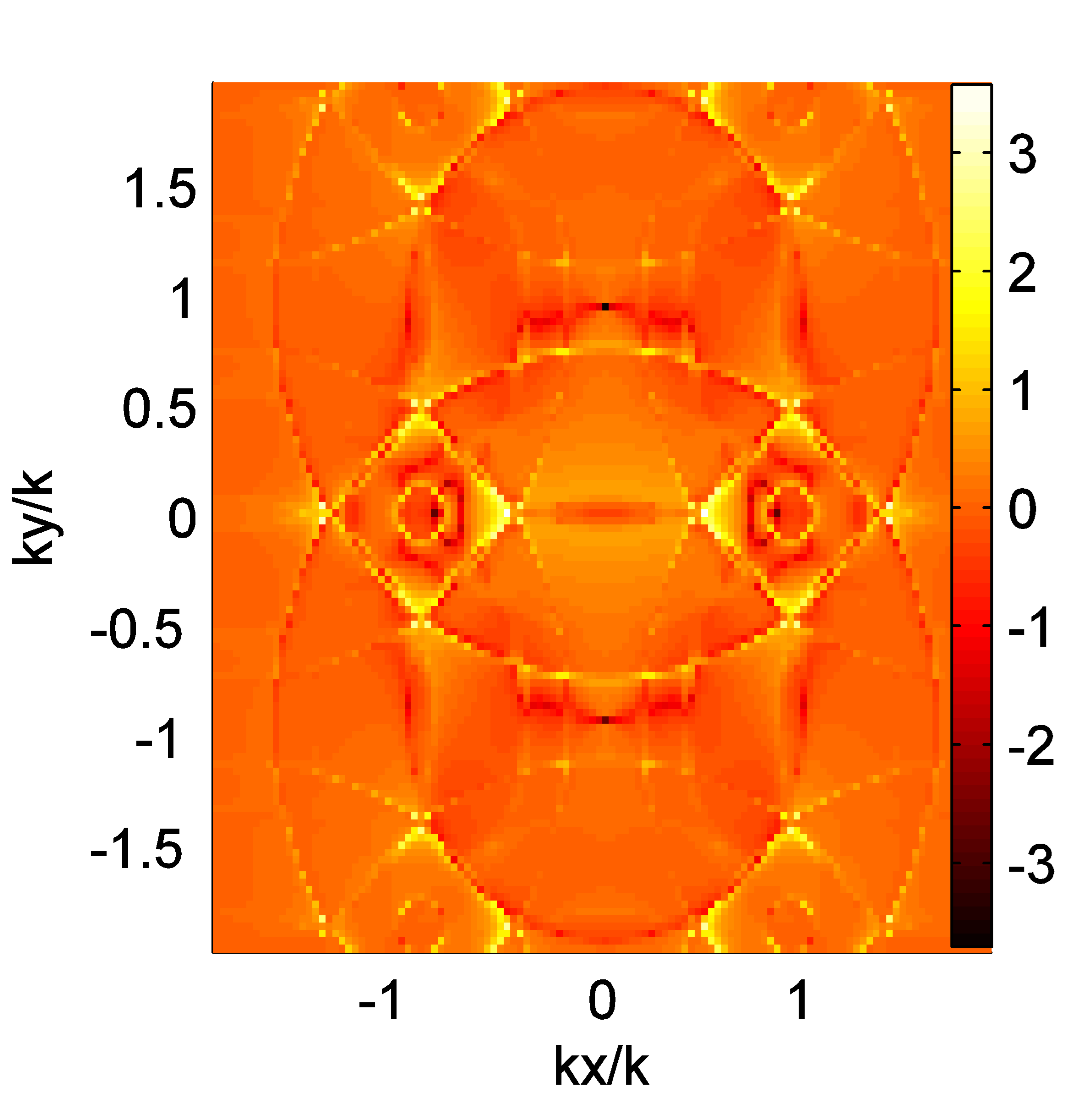} &
\includegraphics[height = 5.5cm]{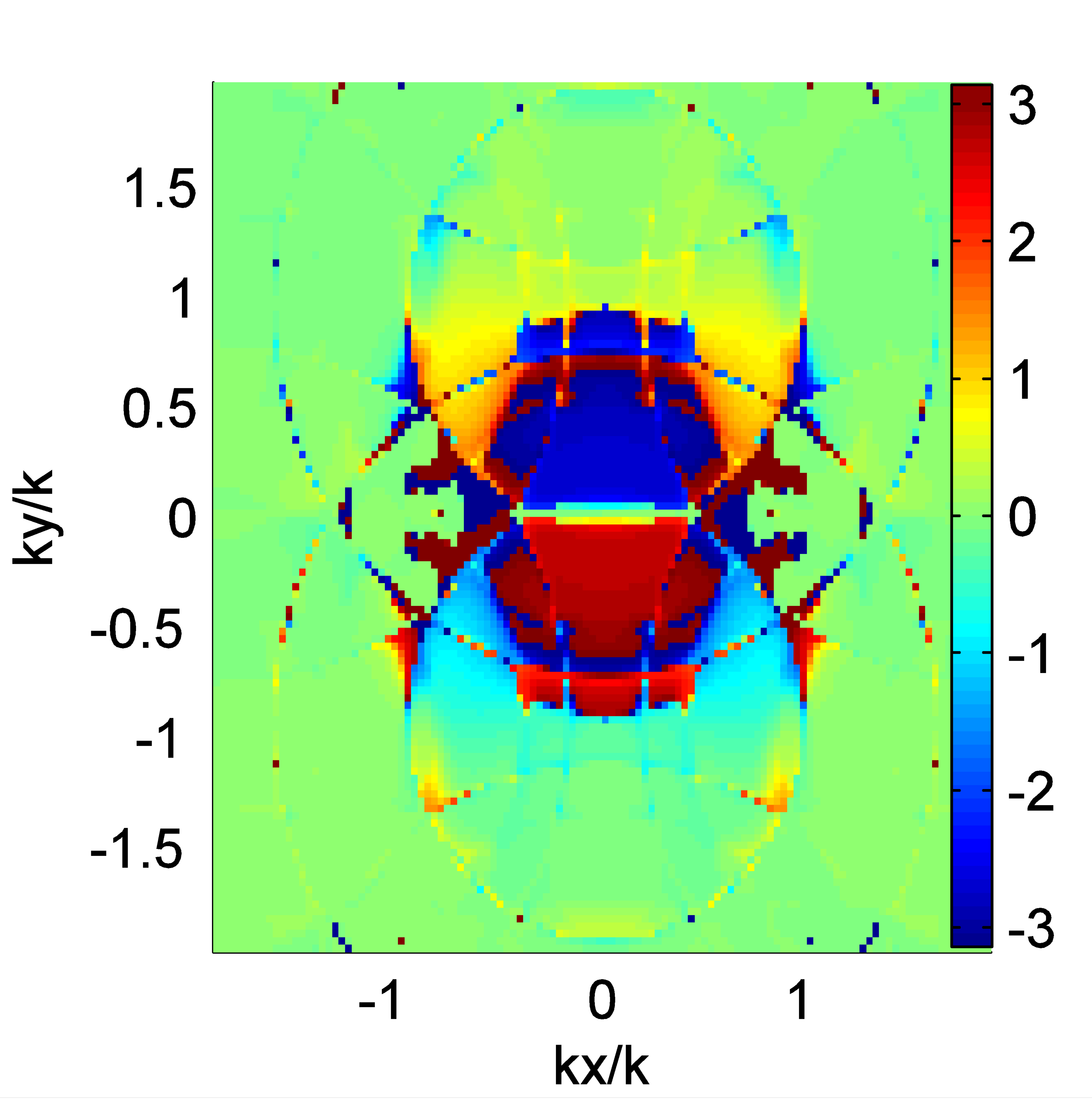} \\
~~~~~~(e) & ~~~~~~(f)
\end{tabular}
\caption{Shape of the $\underline{\underline{\mathcal{H}_{mn}}}(\mathbf{k}_t)$ functions. The logarithm with base 10 of the norm of the function has been represented for better readability. (a) : Norm of entry (1,1) of $\underline{\underline{\mathcal{H}_{00}}}(\mathbf{k}_t)$, which corresponds to the power absorbed for a TE incident wave $\mathbf{E}^{TE}(\mathbf{r}|\mathbf{k}_t)$. (b) : Norm of entry (2,2) of $\underline{\underline{\mathcal{H}_{00}}}(\mathbf{k}_t)$, which corresponds to the power absorbed for a TM incident wave $\mathbf{E}^{TM}(\mathbf{r}|\mathbf{k}_t)$.  (c) and (d): Norm and phase of the entry (1,2) of $\underline{\underline{\mathcal{H}_{00}}}(\mathbf{k}_t)$, which corresponds to the coupling between TE and TM incident waves,  $\mathbf{E}^{TE}(\mathbf{r}|\mathbf{k}_t)$ and $\mathbf{E}^{TM}(\mathbf{r}|\mathbf{k}_t)$ respectively. (e) and (f): Norm and phase of the entry (1,1) of $\underline{\underline{\mathcal{H}_{01}}}(\mathbf{k}_t)$, which corresponds to the coupling between the $\mathbf{E}^{TE}(\mathbf{r}|\mathbf{k}_t-\mathbf{k}_{t}^{01})$ and the $\mathbf{E}^{TE}(\mathbf{r}|\mathbf{k}_t+\mathbf{k}_{t}^{01})$ waves.}
\label{fig:Hmn}
\end{figure*}

First, we computed $\underline{\underline{\mathcal{H}_{mn}}}(\mathbf{k}_t)$ using (\ref{eq:17-09-01}). The results can be seen in Figure \ref{fig:Hmn} for different $m,n$ and different entries of the matrix. The first two graphs having purely real values, their phase is not displayed. The log-magnitude of the field is shown for better readability.

First we consider Fig. \ref{fig:Hmn}(a). Two circles centered on $k_x = k_y = 0$ are visible. The inner one corresponds to the limit of the visibility angle. Inside that circle, modes are propagative in free space while outside modes are evanescent. The $\underline{\underline{\mathcal{H}_{00}}}(\mathbf{k}_t)$ function has to vanish when approaching this circle from the inside, as the power carried by the incident plane wave is vanishing for grazing incidence. The second circle corresponds to a leaky surface mode of the structure (see for example \citep{leaky_wave} and references therein). Most of the other noticeable inhomogeneities correspond to the replicas of these circles into other Brillouin zones \citep{Pozar84}.

Looking at Fig. \ref{fig:Hmn}(b), we can observe similar circles. Interestingly, the wave vector of the surface leaky wave is slightly different for TM or TE excitations ($1.5 k_0$ vs. $1.7k_0$, respectively), corresponding to the fact that there exist two different surface leaky waves. The central square is due to the excitation of this TM leaky wave through its harmonics, while the others inhomogeneities in the visible region are due to the excitation of the TE leaky wave.

The graph (c) in Fig. \ref{fig:Hmn} clearly exhibits the features associated with the visibility limit and the two leaky surface waves. In addition, some straight lines for which the coupling between TE and TM modes vanishes appear due to the symmetries of the structure. Finally, in graph (e), we can see that the circles are centered on $k_y = \pm 0.9375 k_0$. The reciprocal lattice vectors for a periodicity of $400$ nm corresponding to $2 \times 0.9375 k_0$, it is coherent with the expected results. 

\begin{figure}[!ht]
\center
\includegraphics[width = 8.5cm]{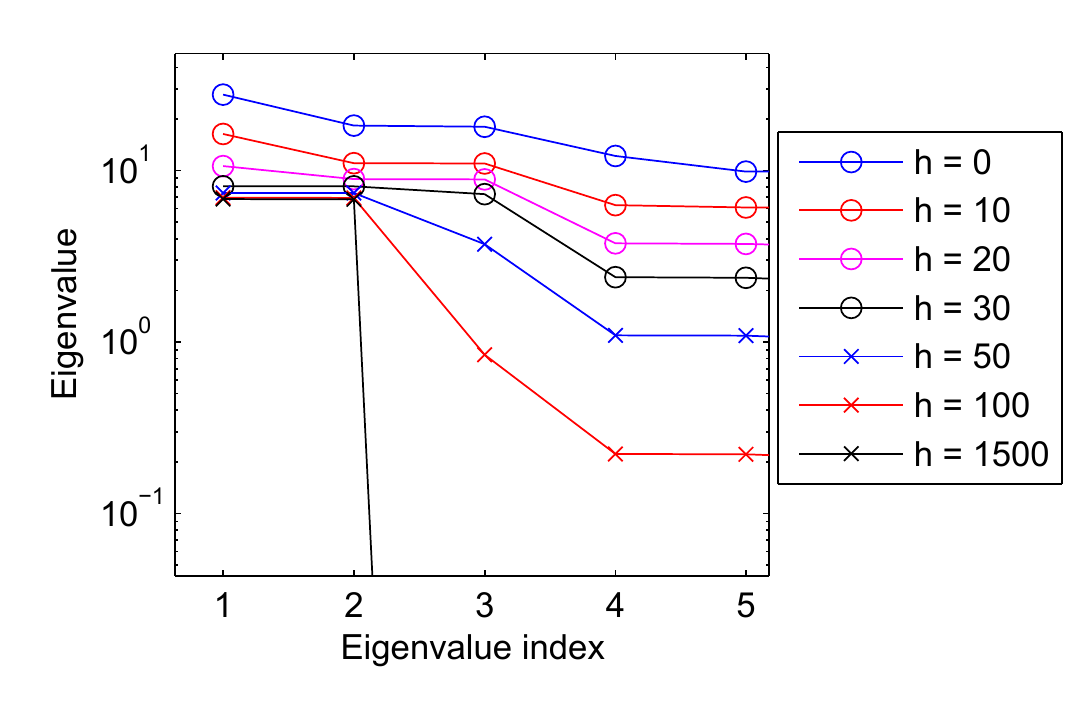}
\caption{Eigenvalues associated to the absorption modes of the structure. The eigenmodes have been computed for different distances between the sources and the structure. The distance is in nm.}
\label{fig:eigenval}
\end{figure}

We also computed the natural absorption modes of the structure and their corresponding eigenvalues for periodic excitations with no phase shift between consecutive unit cells. The eigenvalues were computed for different distances between the reference plane and the absorbing structure. The results can be seen in Figure \ref{fig:eigenval}. We observe that the closer the sources are to the structure, the higher is the number of different modes through which the structure can absorb energy. It is consistent if we consider that evanescent modes can transport energy if there is some reflection on the surface of the structure, but that the amplitude of these modes will decrease over the distance separating the reference plane and the structure. 
Further, we see that, as the distance of the sources from the observer increases, the far-field solution is eventually recovered.

As a final illustration of the technique, we consider the following example. The structure simulated here above is illuminated by an incident plane wave which is partially polarized into TE mode. This plane wave is split into two parts of equal amplitude and one of the two beams goes through a TE polarizer. The incident fields are not perfectly single-frequency and are therefore characterized by a finite coherence length, which is supposed to be much larger than the typical dimensions of one unit-cell of the structure under study. The two beams follow paths with different lengths before reaching the structure, so that the amplitude of the two waves is not perfectly coherent. The non-polarized wave reaching the structure is characterized by a transverse wave vector $\mathbf{k}_t^{(1)} = (-\pi/a, 0, 0)$ and the polarized beam is characterized by the transverse wave vector $\mathbf{k}_t^{(2)} = (\pi/a, 0, 0)$. These two wave vectors correspond to an identical phase shift between consecutive unit cells, so that the power absorbed due to one mode depends on the amplitude of the other mode.  We want to compute the way in which this excitation can be absorbed by the structure.

First, the fields exciting the structure can be characterized using the spectral cross-correlation function $\underline{\underline{\tilde{W}}}(\mathbf{k}_t, \mathbf{k}_t')$. The excitation consisting of only two plane waves, we have
\begin{subequations}
\begin{align}
 \underline{\underline{\tilde{W}}}(\mathbf{k}_t^{(1)}, \mathbf{k}_t^{(1)}) &= 
 \begin{pmatrix} 
 0.5+A & 0 \\
 0 & 0.5-A
 \end{pmatrix}\\
 \underline{\underline{\tilde{W}}}(\mathbf{k}_t^{(1)}, \mathbf{k}_t^{(2)}) &= 
 \begin{pmatrix} 
 B(0.5+A) & 0 \\
 0 & 0
 \end{pmatrix}\\
 \underline{\underline{\tilde{W}}}(\mathbf{k}_t^{(2)}, \mathbf{k}_t^{(2)}) &= 
 \begin{pmatrix} 
 0.5+A & 0 \\
 0 & 0
 \end{pmatrix},
 \end{align}
\end{subequations}
with $A$ a real factor that describes the degree of polarization of the incident wave and $B$ a complex factor that depends on the coherence length of the excitation and the path difference of the two beams. $A=0.5$ corresponds to a perfectly polarized TE wave while $A=0$ corresponds to a perfectly unpolarized wave. If the two beams travel over the same distance, they arrive in phase and $B = 1$. But if the lengths the two beams travel are different, $B$ becomes complex and its amplitude tends to decrease, until the difference between paths is much bigger than the coherence length, in which case $|B| \rightarrow 0$.  Using (\ref{eq:wtilde}), it can be shown that $ \underline{\underline{\tilde{W}}}(\mathbf{k}_t^{(2)}, \mathbf{k}_t^{(1)}) =  \underline{\underline{\tilde{W}}}(\mathbf{k}_t^{(1)}, \mathbf{k}_t^{(2)})^\dagger$. 

We can decompose this correlation matrix into a sum of properly weighted perfectly coherent modes (cf. Equation (\ref{eq:mode_decomp})), which provides

\begin{subequations}
\begin{align}
\mathbf{\tilde{E}}_1(\mathbf{k}_t^{(1)}) = 
	\begin{pmatrix}
	0\\1
	\end{pmatrix} ~~~~
	\mathbf{\tilde{E}}_1&(\mathbf{k}_t^{(2)}) = 
	\begin{pmatrix}
	0 \\ 0
	\end{pmatrix} \nonumber\\
&\lambda_1 = 0.5-A 
\end{align}
\begin{align}
\mathbf{\tilde{E}}_2(\mathbf{k}_t^{(1)}) = \dfrac{1}{\sqrt{2}}
	\begin{pmatrix}
	1\\0
	\end{pmatrix} ~~~~
	\mathbf{\tilde{E}}_2&(\mathbf{k}_t^{(2)}) = \dfrac{1}{\sqrt{2}}
	\begin{pmatrix}
	B/|B| \\ 0
	\end{pmatrix}\nonumber\\
&\lambda_2 = (1+|B|)(0.5+A) 
\end{align}
\begin{align}
\mathbf{\tilde{E}}_3(\mathbf{k}_t^{(1)}) = \dfrac{1}{\sqrt{2}} 
	\begin{pmatrix}
	1\\0
	\end{pmatrix} ~~~~
	\mathbf{\tilde{E}}_3&(\mathbf{k}_t^{(2)}) = \dfrac{1}{\sqrt{2}} 
	\begin{pmatrix}
	-B/|B| \\ 0
	\end{pmatrix}\nonumber\\
&\lambda_3 = (1-|B|)(0.5+A) 
\end{align}
\end{subequations}

From a physical point of view, we can see that the partially coherent excitation can be decomposed into 
\begin{enumerate}
\item a TM plane wave of transverse wave vector $\mathbf{k}_t^{(1)}$,
\item two TE plane waves of equal amplitude and transverse vectors $\mathbf{k}_t^{(1)}$ and $\mathbf{k}_t^{(2)}$,
\item two TE plane waves of opposite amplitude and transverse vectors $\mathbf{k}_t^{(1)}$ and $\mathbf{k}_t^{(2)}$.
\end{enumerate}
The amplitude of each of these modes depends on the degree of coherence of the incident fields. The total power dissipated inside the structure will correspond to the sum of the powers absorbed by each of these modes separately. 

Alternatively, we can compute the natural modes of absorption of the structure from the data computed previously (see Fig. \ref{fig:Hmn}). Since we are dealing with a far-field excitation, only four terms have to be taken into account in Equation (\ref{eq:18-12-01}), whose associated fields distribution are a linear combination of the four propagating Floquet modes associated to the $\pi$ phase shift between consecutive unit cells.

In matrix form, the eigenmodes and the associated eigenvalues are 
\begin{subequations}
\begin{align}
\mathcal{E}_1(\mathbf{k}_t^{(1)}) = \dfrac{1}{\sqrt{2}}
	\begin{pmatrix}
	0\\1
	\end{pmatrix} ~~~~
	\mathcal{E}_1 & (\mathbf{k}_t^{(2)}) = \dfrac{1}{\sqrt{2}}
	\begin{pmatrix}
	0 \\ -1
	\end{pmatrix} \nonumber \\
&\Lambda_1 = 5.75 \\
~ \nonumber \\
\mathcal{E}_2(\mathbf{k}_t^{(1)}) = \dfrac{1}{\sqrt{2}}
	\begin{pmatrix}
	1\\0
	\end{pmatrix} ~~~~
	\mathcal{E}_2 & (\mathbf{k}_t^{(2)}) = \dfrac{1}{\sqrt{2}}
	\begin{pmatrix}
	1\\0
	\end{pmatrix}\nonumber\\
&\Lambda_2 = 3.58 
\end{align}
\begin{align}
\mathcal{E}_3(\mathbf{k}_t^{(1)}) = \dfrac{1}{\sqrt{2}}
	\begin{pmatrix}
	1\\0
	\end{pmatrix} ~~~~
	\mathcal{E}_3 & (\mathbf{k}_t^{(2)}) = \dfrac{1}{\sqrt{2}}
	\begin{pmatrix}
	-1\\0
	\end{pmatrix}\nonumber\\
&\Lambda_3 = 2.74 \\
~ \nonumber \\
\mathcal{E}_4(\mathbf{k}_t^{(1)}) = \dfrac{1}{\sqrt{2}} 
	\begin{pmatrix}
	0\\1
	\end{pmatrix} ~~~~
	\mathcal{E}_4 & (\mathbf{k}_t^{(2)}) = \dfrac{1}{\sqrt{2}} 
	\begin{pmatrix}
	0\\1
	\end{pmatrix}\nonumber\\
&\Lambda_4 = 2.06  .
\end{align}
\end{subequations}
Applying Equation (\ref{eq:15-01-04}), we can see that the total power dissipated by each unit cell of the structure (in arbitrary units) is
\begin{equation}
\begin{split}
P = 3.905 &(0.5-A) \\
+ (1&+|B|)  (0.5+A) \\
\times &\Big( 0.895 \big|1+B/|B|\big|^2 +0.685 \big|1-B/|B|\big|^2 \Big) \\
+ (1&-|B|)  (0.5+A) \\
\times &\Big( 0.895 \big|1-B/|B|\big|^2   +0.685 \big|1+B/|B|\big|^2 \Big) .
\end{split}
\end{equation}

As we can see, provided that the absorbing modes of the structure are known, a closed form expression of the power absorbed by the structure when it is illuminated by any partially coherent fields can be easily derived.

\section{Conclusion}
When computing the absorption of partially spatially coherent and partially polarized incident fields by a given structure, fully coherent calculations fail to deal with the stochastic nature of the fields. We have shown that the problem can be carried out in two different ways. In the first scheme, the spectral cross-correlation function $\underline{\underline{W}}(\mathbf{k}_t, \mathbf{k}_t')$ of the incident field is decomposed into a set of fully spatially coherent and polarized modes, each of which is excited incoherently with respect to the others. Since these modes are individually fully spatially coherent, their interaction with any structure can be treated using coherent-field solvers. In the second scheme, the orthogonal natural absorption modes of the structure are computed. The total power absorbed for any incident fields is obtained by the projection of the spectral cross-correlation function of the incident fields on these modes.

The natural absorption modes of a structure can be computed using the cross-spectral power density function $\underline{\underline{P_o}}(\mathbf{k}_t, \mathbf{k}_t')$ of the structure, which fully characterizes the way the structure can absorb any incident fields at a given frequency. This function can be obtained both numerically and experimentally, using the Energy Absorption Interferometry already described in the literature. The 4D cross-spectral power density function can be represented using a discrete compact set of 2D functions $\underline{\underline{H_{mn}}}(\mathbf{k}_t)$ in the case where the absorbing structure is 2D periodic. We also provided some recommendation for truncating these functions.

The formulation presented in this paper is general as it can deal with real 3D geometries and with the polarization of the incident fields. Both numerical and analytical examples are provided to assess its efficiency.

\section*{Acknowledgments}

Computational resources have been provided by the supercomputing facilities of the Universit\'{e} catholique de Louvain (CISM/UCL) and the Consortium des Equipements de Calcul Intensif en F\'{e}d\'{e}ration Wallonie Bruxelles (CECI) funded by the Fond de la Recherche Scientifique de Belgique (F.R.S.-FNRS) under convention 2.5020.11.

\end{document}